\documentclass[12pt]{iopart}

\usepackage{mathrsfs}
\usepackage{epsfig}
\usepackage{graphicx}

\begin{document}

\title{Dynamical Decoupling in Common Environment}

\author{Yu Pan, Hong-Ting Song, Zai-Rong Xi}
\address{Key Laboratory of Systems and
Control, Institute of Systems Science, Academy of Mathematics and
Systems Science, Chinese Academy of Sciences, Beijing 100190,
People's Republic of China} \ead{zrxi@iss.ac.cn}

\begin{abstract}
Dynamical decoupling (DD) sequences were invented to eliminate the
direct coupling between qubit and its environment. We further
investigate the possibility of decoupling the indirect qubit-qubit
interaction induced by a common environment, and sucessfully find
simplified solutions that preserve the bipartite quantum states to
arbitrary order. Through analyzing the exact dynamics of the
controlled two-qubit density matrix, we have proven that applying
independent Uhrig Dynamical Decoupling (UDD) on each qubit will
effectively eliminate both the qubit-environment and indirect
qubit-qubit coupling to the same order as in single qubit case, only
if orders of the two UDD sequences have different parity. More
specifically, UDD($n$) on one qubit with UDD($m$) on another are
able to produce $\min(n,m)$th order suppression while $n+m$ is odd.
Our results can be used to reduce the pulse number in relevant
experiments for protecting bipartite quantum states, or dynamically
manipulate the indirect interaction within certain quantum gate and
quantum bus.

\end{abstract}

\maketitle

\section{Introduction}

Qubit is constantly losing its coherence due to interaction with
environment. Dynamical decoupling sequences are proposed to
eliminate the unwanted qubit-environment couplings with
instantaneous $\pi$ pulses, and efficiency of DD sequences is
largely dependent on the pulse locations. Equidistant DD is a first
order solution \cite{ar1}, while more advanced sequences employ
non-equidistant \cite{ar2,ar3} or concatenated pulse locations
\cite{ar4,ar5}. Particularly in this article we are concerned about
UDD \cite{ar2}, whose pulse locations are
$t\delta_j$$$\delta_j=\sin^2(j\pi/2(n+1)).$$UDD sequence was first
derived under pure dephasing models \cite{ar6}, able to protect the
qubit coherence up to $n$th order with $n$ pulses.\par In recent
years, DD has been successfully extended to preserve multipartite
quantum states \cite{ar7,ar8,ar9,ar10,ar11,ar12,ar13}. By nesting
layers of UDD (NUDD), total $(n+1)^m$ pulses are needed to freeze a
system of $m$ qubits to $n$th order without knowing any details of
the qubit-environment coupling \cite{ar8,ar10,ar14,ar15,ar16}. It
can be seen that in NUDD pulse number grows polynomial with pulse
order. However for a specific experimental setup, it is more
realistic and efficient to tailor the pulse sequence accordingly,
since pulse number is limited during a fixed time interval
\cite{ar17,ar18} and errors from each imperfect pulse will
accumulate \cite{ar19,ar20,ar21}. There are already some efforts to
reduce the pulse number. If prior knowledge is available, such as
initial qubit state \cite{ar7}, environmental coupling spectrum
\cite{ar11} or unbalanced decoherence rates \cite{ar12}, the above
universal plan can be greatly simplified.\par In this paper we
introduce another approach to reduce the DD sequence which is needed
for a two-qubit system linearly coupled to a common bosonic bath.
This situation may arise when two qubits are not spatially separated
enough to create independent environments \cite{ar22}, or noises
felt by each qubit are correlated \cite{ar11,ar23}. In addition,
common environment has long been exploited to generate entangling
gate and serve as quantum bus between qubits
\cite{ar24,ar25,ar26,ar27}. While these gates and buses are idle we
can use DD sequences to switch off the interactions and in the
meantime protect the states.\par As it turns out, the simplified
sequence is made very easy to implement. For example, we can apply
UDD($n$) ($n$-pulse UDD) on first qubit and UDD($m$) on second
qubit. $n$ and $m$ can be chosen at will as long as they have
different parity. The whole sequence will preserve the bipartite
state up to an order of $\min(n,m)$. For $m=n+1$, this scheme only
needs $2n+1$ pulses, while NUDD($n$) needs $(n+1)^2$.\par We
organize this paper as follows. In Sec. II we discuss the free
dynamics of a two-qubit system in a common bosonic bath. In Sec.
III, controlled dynamics and conditions for high-level DD are
derived. Sequences satisfying these conditions are given.
Conclusions are put in Sec. IV.

\section{Dynamics in Common Quantum Bath}

We consider a two-qubit spin-boson model:
\begin{eqnarray}
H=\sum_k\omega_kb_k^{\dagger}b_k+(\sigma_{z_1}+\sigma_{z_2})\sum_k\lambda_k(b_k^{\dagger}+b_k),
\end{eqnarray}
where $b_k$ is the annihilation operator for the $k$th mode of the
bath and $\lambda_k$ is coupling strength. $\sigma_{z_1}$ and
$\sigma_{z_2}$ are spin operators acting on first and second qubit
respectively. This model describes a pure dephasing process due to
the couplings with environment. With $|\uparrow\rangle$ and
$|\downarrow\rangle$ being the eigenstates of $\sigma_z$, we can
define four basis states $|0\rangle=|\downarrow\downarrow\rangle,
|1\rangle=|\downarrow\uparrow\rangle,
|2\rangle=|\uparrow\downarrow\rangle,
|3\rangle=|\uparrow\uparrow\rangle$ for a bipartite quantum
state.\par The dynamics of Eq.~($1$) can be exactly solved
\cite{ar25,ar28}. In the interaction picture of the bath operator
$\sum_k\omega_kb_k^{\dagger}b_k$, the unitary evolution that
generated by the time-dependent interaction Hamiltonian can be
computed using Magnus expansion. It is easy to verify that only the
first two terms of the expansion are nonzero (see \cite{ar28} for
more details). As a result, the free evolution of the composite
system can be calculated as follows
\begin{eqnarray}
U_f(t)&=&\exp[(\sigma_{z_1}+\sigma_{z_2})\sum_k\lambda_k(\frac{e^{-\mbox{i}\omega_kt}-1}{\omega_k}b_k-\frac{e^{\mbox{i}\omega_kt}-1}{\omega_k}b_k^\dagger)\nonumber\\
&+&\mbox{i}(\sigma_{z_1}+\sigma_{z_2})^2\sum_k\lambda_k^2\frac{\omega_kt-\sin\omega_kt}{\omega_k^2}].
\end{eqnarray}
The collective term
$(\sigma_{z_1}+\sigma_{z_2})^2=2(I+\sigma_{z_1}\sigma_{z_2})$
generates an indirect coupling between the two qubits dependent on
their states, which will further induce an oscillation of quantum
correlations \cite{ar25,ar27,ar29}. In order to preserve an
arbitrary state, not only couplings to bath oscillators but also the
indirect couplings must be removed. By introducing the spectral
density function \cite{ar30} defined as
$J(\omega)=\sum_k\lambda_k^2\delta(\omega-\omega_k)$, we can then
write down the evolution of the density matrix $\rho(t)$
$$\left(
  \begin{array}{cccc}
    1 & {e^{\mbox{i}\Delta(t)-\Gamma(t)}} & {e^{\mbox{i}\Delta(t)-\Gamma(t)}} & {e^{-4\Gamma(t)}}\\
    {e^{-\mbox{i}\Delta(t)-\Gamma(t)}} & 1 & 1 & {e^{-\mbox{i}\Delta(t)-\Gamma(t)}}\\
    {e^{-\mbox{i}\Delta(t)-\Gamma(t)}} & 1 & 1 & {e^{-\mbox{i}\Delta(t)-\Gamma(t)}}\\
    {e^{-4\Gamma(t)}} & {e^{\mbox{i}\Delta(t)-\Gamma(t)}} & {e^{\mbox{i}\Delta(t)-\Gamma(t)}} & 1
  \end{array}
\right),$$ with
\begin{eqnarray}
\Delta(t)&=&4\int_0^{\infty}J(\omega)\frac{\omega{t}-\sin\omega{t}}{\omega^2}d\omega,\\
\Gamma(t)&=&4\int_0^{\infty}J(\omega)\frac{2\sin^2(\omega{t}/2)}{\omega^2}\coth(\frac{\beta\omega}{2})d\omega.
\end{eqnarray}
$\beta$ is the inverse temperature. The exponentially decaying
factor $\Gamma(t)$ is associated with decoherence process, while
$\Delta(t)$ represents collective phase evolution. Note that the
phase factor is absent in classical noise model \cite{ar11,ar23}. If
two qubits are subjected to the same classical noise, it is adequate
to use simultaneous DD pulses on each qubit.\par In the case of
quantum noise, still by flipping the sign of $\sigma_{z_1}$ and
$\sigma_{z_2}$ with $\pi$ pulses, $\Gamma(t)$ will be effectively
averaged to zero. However, we cannot apply the same sequence on both
qubits, since simultaneous $\pi$ pulses will have no influence on
the value of $(\sigma_{z_1}+\sigma_{z_2})^2$ and so on phase
evolution. This motivates us to consider the following scenario: $n$
pulses of $\pi$ rotation along axis-$x$ are applied to the first
qubit, with the pulse locations given by $\delta_{1^{'}},
\delta_{2^{'}},..., \delta_{n^{'}}$, whereas $m$ analogous pulses
are applied to the second qubit at different times $\delta_{1^{''}},
\delta_{2^{''}},..., \delta_{{m}^{''}}$. Totally $n+m$ pulses are
applied to the two-qubit system. We arrange the $n+m$ pulse timings
in increasing order and denote them by $\delta_1, \delta_2,...,
\delta_{n+m}$, with $\delta_j<\delta_{j+1}$. At each $\delta_{j}$,
either the $\sigma_{z_1}$ or $\sigma_{z_2}$ operator switches its
sign. At the same time, the operator $\sigma_{z_1}\sigma_{z_2}$
changes its sign $n+m$ times at these instants. In the next section
we seek to find the correct $n+m$ pulse locations which preserve
both phases and amplitudes of the density matrix elements.

\section{Pulse Controlled Dynamics}

In this section we adopt the canonical transformation technique
which was used in deriving the UDD sequence for single qubit
\cite{ar2,ar6}. Also we use the notations from \cite{ar6} by
defining
\begin{eqnarray}
A^{\mbox{eff}}&=&UAU^{\dagger},A(t)=\exp(\mbox{i}H^{\mbox{eff}}t)A\exp(-\mbox{i}H^{\mbox{eff}}t),\\
U&=&\exp[(\sigma_{z_1}+\sigma_{z_2})K],K=\sum_k\frac{\lambda_k}{\omega_k}(b_k^\dagger-b_k).
\end{eqnarray}
Operator $A^{\mbox{eff}}$ acquires time dependence under the action
of the ``effective" Hamiltonian. After the canonical transformation
$U$, the effective Hamiltonian is diagonal
\begin{eqnarray}
H^{\mbox{eff}}=\sum_k\omega_kb_k^{\dagger}b_k-\frac{\lambda_k^2}{\omega_k}(\sigma_{z_1}+\sigma_{z_2})^2
\end{eqnarray}
and the time-dependent flip operators $\sigma_{x_i}^{\mbox{eff}}(t)$
are
\begin{eqnarray}
\sigma_{x_i}^{\mbox{eff}}(t)=\exp[2\sigma_{z_i}K(t)]\exp(-4\mbox{i}\sum_k\frac{\lambda_k^2}{\omega_k}\sigma_{z_1}\sigma_{z_2}t)\sigma_{x_i},
\end{eqnarray}
with
\begin{eqnarray}
K(t)=\sum_k\frac{\lambda_k}{\omega_k}(b_k^{\dagger}e^{\mbox{i}\omega_kt}-b_ke^{-\mbox{i}\omega_kt}).
\end{eqnarray}
Making use of these expressions, dynamics of the density matrix
elements can be obtained by explicit calculation. For an arbitrary
element $\rho_{SS^{'}}$, the average evolution is
\begin{eqnarray}
&\langle{\langle}S|e^{\mbox{i}Ht}|S\rangle{\langle}S^{'}|e^{-\mbox{i}Ht}|S^{'}\rangle\rangle&\nonumber\\
&=\langle{\langle}S|e^{\mbox{i}H(\delta_{1}-\delta_{0})t}\sigma_{x_{j_1}}e^{\mbox{i}H(\delta_{2}-\delta_{1})t}...e^{\mbox{i}H(\delta_{n+m}-\delta_{n+m-1})t}\sigma_{x_{j_{n+m}}}e^{\mbox{i}H(\delta_{n+m+1}-\delta_{n+m})t}(\sigma_{x_i})|S\rangle&\nonumber\\
&{\langle}S^{'}|(\sigma_{x_i})e^{-\mbox{i}H(\delta_{n+m+1}-\delta_{n+m})t}\sigma_{x_{j_{n+m}}}e^{-\mbox{i}H(\delta_{n+m}-\delta_{n+m-1})t}...e^{-\mbox{i}H(\delta_{2}-\delta_{1})t}\sigma_{x_{j_{1}}}e^{-\mbox{i}H(\delta_{1}-\delta_{0})t}|S^{'}\rangle\rangle&\nonumber\\
&=\langle{\langle}S|U^{\dagger}\sigma_{x_{j_{1}}}(\delta_{1}t)\sigma_{x_{j_{2}}}(\delta_{2}t)...\sigma_{x_{j_{n+m-1}}}(\delta_{n+m-1}t)\sigma_{x_{j_{n+m}}}(\delta_{n+m}t)(\sigma_{x_i}^{\mbox{eff}}(t))e^{\mbox{i}H^{\mbox{eff}}t}U|S\rangle\nonumber&\\
&{\langle}S^{'}|U^{\dagger}e^{-\mbox{i}H^{\mbox{eff}}t}(\sigma_{x_i}^{\mbox{eff}}(t))\sigma_{x_{j_{n+m}}}(\delta_{n+m}t)\sigma_{x_{j_{n+m-1}}}(\delta_{n+m-1}t)...\sigma_{x_{j_{2}}}(\delta_{2}t)\sigma_{x_{j_{1}}}(\delta_{1}t)U|S^{'}\rangle\rangle&\nonumber\\
&=\langle{\langle}S|U^{\dagger}\sigma_{x_{j_{1}}}(\delta_{1}t)\sigma_{x_{j_{2}}}(\delta_{2}t)...\sigma_{x_{j_{n+m-1}}}(\delta_{n+m-1}t)\sigma_{x_{j_{n+m}}}(\delta_{n+m}t)(\sigma_{x_i}^{\mbox{eff}}(t))&\nonumber\\
&e^{(s_1-s_2)K(t)}e^{-\mbox{i}\sum_k\frac{\lambda_k^2}{\omega_k}({s_1}^2-{s_2}^2)t}|S\rangle{\langle}S^{'}|(\sigma_{x_i}^{\mbox{eff}}(t))\sigma_{x_{j_{n+m}}}(\delta_{n+m}t)&\nonumber\\
&\sigma_{x_{j_{n+m-1}}}(\delta_{n+m-1}t)...\sigma_{x_{j_{2}}}(\delta_{2}t)\sigma_{x_{j_{1}}}(\delta_{1}t)U|S^{'}\rangle\rangle.&
\end{eqnarray}
$s_1,s_2$ are eigenvalues of $\sigma_{z_1}+\sigma_{z_2}$ for
$|S\rangle$ and $|S^{'}\rangle$. Determined by which spin is flipped
at $\delta_j$, $j_1,j_2,...j_{n+m}$ take values from $\{1,2\}$. By
$(\sigma_{x_i})$ we mean that an ending pulse $\sigma_{x_i}$ is
added at the end of the sequence to ensure the final state is still
$|S\rangle{\langle}S^{'}|$ if $n+m$ is odd (If $n$ and $m$ are both
odd, two ending pulses are needed). However according to our
calculation, whether or not there is an (or two) ending pulse will
not modify the equations and results we derive next. The outermost
brackets denote ensemble average with respect to the thermal
bath.\par The sum of all terms in the form of
$-4\mbox{i}\sum_k(\lambda_k^2/\omega_k)\sigma_{z_1}\sigma_{z_2}$
from Eq.~($10$) leads to a phase shift
\begin{eqnarray}
\exp[-4\mbox{i}\sum_k\frac{\lambda_k^2}{\omega_k}(s_1-s_2)(\delta_1-\delta_2+...+(-1)^{n+m-1}\delta_{n+m}+\frac{(-1)^{n+m}}{2})t],\nonumber\\
\end{eqnarray}
and calculation of the rest part of Eq.~($10$) will produce more
phases. Taking $\rho_{01}(t)$ as an example, we are left with
\begin{eqnarray}
{\langle}e^{2K}e^{-2K(\delta_{1}t)}e^{(\pm)2K(\delta_{2}t)}...e^{(\pm)2K(\delta_{m+n}t)}e^{(-1)^{m+1}2K(t)}\nonumber\\
e^{(\pm)2K(\delta_{m+n}t)}...e^{(\pm)2K(\delta_{2}t)}e^{-2K(\delta_{1}t)}\rangle.\nonumber\\
\end{eqnarray}
Coefficients before each $K(\delta_it)$ are determined by the
relative locations between 1st-qubit and 2nd-qubit pulses in the
whole sequence. Exponentials in Eq.~($12$) can be combined one by
one using Baker-Campbell-Hausdorff formula
$e^Ae^B=e^{A+B}e^{[A,B]/2}$, which is valid here because
\begin{eqnarray*}
[K(t),K(t^{'})]=2\mbox{i}\sum_k(\lambda_k^2/\omega_k^2)\sin\omega_k(t-t^{'})
\end{eqnarray*}
is a $c$-number. After the combination all $K(\delta_it)$ add up to
one exponential
$\exp[-2K(\delta_{1^{''}}t)+2K(\delta_{2^{''}}t)...+(-1)^m2K(\delta_{m^{''}}t)+(-1)^{m+1}2K(t)+(-1)^m2K(\delta_{m^{''}}t)...+2K(\delta_{2^{''}}t)-2K(\delta_{1^{''}}t)]$
while extra phases introduced by the $e^{[A,B]/2}$ term in Hausdorff
formula can be calculated by observing that combining any four
symmetric exponentials $e^{\pm2K(\delta_pt)}$ and
$e^{\pm2K(\delta_qt)}$ on both sides of $e^{(-1)^{m+1}2K(t)}$ will
not create extra phases except for one case: that is when combine
$e^{(-1)^i2K(\delta_{i^{'}}t)}$ with the exponentials which act on
the 2nd qubit after it
\begin{eqnarray}
&e^{(-1)^i2K(\delta_{i^{'}}t)}e^{(-1)^j2K(\delta_{j^{''}}t)}e^{(-1)^j2K(\delta_{j^{''}}t)}e^{(-1)^i(-2)K(\delta_{i^{'}}t)}&\nonumber\\
&=e^{(-1)^{i+j}16\mbox{i}\sum_k\frac{\lambda_k^2}{\omega_k^2}\sin\omega_k(\delta_{i^{'}}-\delta_{j^{''}})t}e^{(-1)^j4K(\delta_{j^{''}}t)},(i^{'}<j^{''}).&\nonumber\\
\end{eqnarray}
Merging these exponentials four by four, we arrive at
\begin{eqnarray}
&\exp\{\mbox{i}\sum_k\frac{\lambda_k^2}{\omega_k^2}[8\sum_{i=1}^n(-1)^{i+m+1}\sin\omega_k(\delta_{i^{'}}-1)t&\nonumber\\
&+16\sum_{i=1}^n\sum_{i^{'}<j^{''}}^m(-1)^{i+j}\sin\omega_k(\delta_{i^{'}}-\delta_{j^{''}})t]\}&\nonumber\\
&{\langle}\exp(2K)\exp[-2K(\delta_{1^{''}}t)+2K(\delta_{2^{''}}t)...&\nonumber\\
&+(-1)^m2K(\delta_{m^{''}}t)+(-1)^{m+1}2K(t)+(-1)^m2K(\delta_{m^{''}}t)&\nonumber\\
&...+2K(\delta_{2^{''}}t)-2K(\delta_{1^{''}}t)]\rangle.&
\end{eqnarray}
The ensemble average in Eq.~($14$) has already been calculated in
\cite{ar6}. With the existing results from \cite{ar6} and combining
Eq.~($11$) and Eq.~($14$), the final form of $\rho_{01}(t)$ can be
organized in a rather compact way
\begin{eqnarray}
\rho_{01}(t)=\exp(\mbox{i}\Delta_m(t)-\Gamma_m(t)),
\end{eqnarray}
with
\begin{eqnarray}
\Delta_m(t)&=&4\int_0^{\infty}\frac{J(\omega)}{\omega^2}(x_m+z_m+c)(\omega{t})d\omega,\nonumber\\
\Gamma_m(t)&=&2\int_0^{\infty}J(\omega)\frac{|y_m(\omega{t})|^2}{\omega^2}\coth(\beta\omega/2)d\omega,\nonumber\\
x_m(\omega{t})&=&(-1)^m\sin(\omega{t})+2\sum_{j=1}^m(-1)^{j+1}\sin(\omega{t}\delta_{j^{''}}),\nonumber\\
y_m(\omega{t})&=&1+(-1)^{m+1}e^{\mbox{i}\omega{t}}+2\sum_{j=1}^m(-1)^je^{\mbox{i}\omega{t}\delta_{j^{''}}},\nonumber\\
z_m(\omega{t})&=&2\sum_{i=1}^n(-1)^{i+m+1}\sin\omega{t}(\delta_{i^{'}}-1)+4\sum_{i=1}^n\sum_{i^{'}<j^{''}}^m(-1)^{i+j}\sin\omega{t}(\delta_{i^{'}}-\delta_{j^{''}}),\nonumber\\
c(\omega{t})&=&-2\omega{t}[\sum_r^{n+m}(-1)^{r-1}\delta_r+\frac{(-1)^{n+m}}{2}].
\end{eqnarray}
Unlike single qubit DD, here the phase factor $\Delta_m(t)$ will
induce collective dynamics and should be minimized. $y_m(\omega{t})$
is the so-called filter function of an $m$-pulse sequence
\cite{ar2,ar3} and $\Gamma_m(t)$ is responsible for decoherence.\par
The most simple way to realize dynamical decoupling is to use
independent UDD sequences to suppress the decoherence, and hope that
phase evolutions will be minimized automatically at the same time.
UDD($m$) requires the first $m$ derivatives of the filter function
to be zero at $\omega=0$. These constraints are imposed to engineer
$y_m(\omega{t})$ in low frequency region, achieving an error order
of $O(t^{m})$. Similarly we can also define the filter function for
$\Delta_m(t)$ as
\begin{eqnarray}
f_m(\omega{t})=x_m(\omega{t})+z_m(\omega{t})+c(\omega{t}).
\end{eqnarray}
The density matrix $\rho(t)$ that controlled by an arbitrary pulse
sequence reads
$$\left(
  \begin{array}{cccc}
    1 & {e^{\mbox{i}\Delta_m(t)-\Gamma_m(t)}} & {e^{\mbox{i}\Delta_n(t)-\Gamma_n(t)}} & {e^{-\Gamma_n(t)-\Gamma_m(t)-\gamma(t)}}\\
    {e^{-\mbox{i}\Delta_m(t)-\Gamma_m(t)}} & 1 & {e^{-\Gamma_n(t)-\Gamma_m(t)-\gamma(t)}} & {e^{-\mbox{i}\Delta_n(t)-\Gamma_n(t)}}\\
    {e^{-\mbox{i}\Delta_n(t)-\Gamma_n(t)}} & {e^{-\Gamma_n(t)-\Gamma_m(t)-\gamma(t)}} & 1 & {e^{-\mbox{i}\Delta_m(t)-\Gamma_m(t)}}\\
    {e^{-\Gamma_n(t)-\Gamma_m(t)-\gamma(t)}} & {e^{\mbox{i}\Delta_n(t)-\Gamma_n(t)}} & {e^{\mbox{i}\Delta_m(t)-\Gamma_m(t)}} & 1
  \end{array}
\right),$$
with
\begin{eqnarray}
\gamma(t)=4\int_0^{\infty}\frac{J(\omega)}{\omega^2}\Re(y_n(\omega{t})y_m^*(\omega{t}))\coth(\beta\omega/2)d\omega.
\end{eqnarray}
Definitions of $\Delta_n(t)$ and $\Gamma_n(t)$ are the same as
Eq.~($16$), but with pulse locations swapped. For example,
\begin{eqnarray}
z_n(\omega{t})&=&2\sum_{j=1}^m(-1)^{j+n+1}\sin\omega{t}(\delta_{j^{''}}-1)\nonumber\\
&+&4\sum_{j=1}^m\sum_{j^{''}<i^{'}}^n(-1)^{i+j}\sin\omega{t}(\delta_{j^{''}}-\delta_{i^{'}}).
\end{eqnarray}
$\rho_{12}$ and $\rho_{21}$ are driven out of the decoherence-free
subspace. In spite of this, it is clear that the exponential decay
of all density matrix elements are filtered by $y_m(\omega{t})$ and
$y_n(\omega{t})$, which is at the order of $\min(n,m)$ if UDD($n$)
and UDD($m$) are applied on each qubit.

\subsection{UDD Sequences with
Different Parity}
\begin{figure}
\begin{center}
\scalebox{0.6}[0.6]{\includegraphics{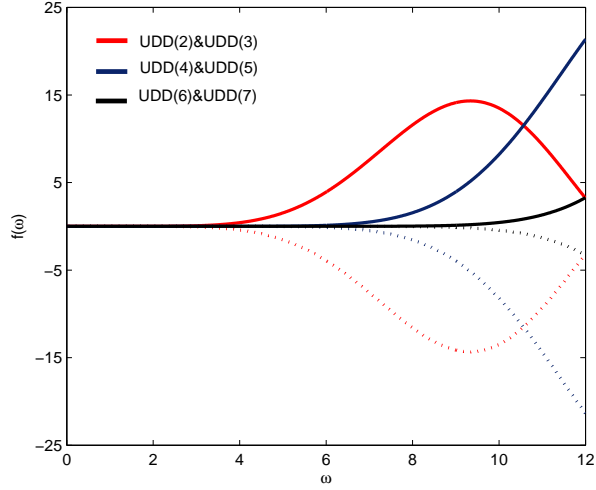}}\caption{$t=1$. Red
line: UDD(2) on the first qubit, UDD(3) on the second qubit. Solid
line for $f_2$, and dashed line for $f_3$. Other four colored lines
follow similar definitions.{\label{fig1}}}
\end{center}
\end{figure}

By requiring the phase shift from Eq.~($11$) vanish, we get the
equality
\begin{eqnarray}
\delta_1-\delta_2+...(-1)^{n+m-1}\delta_{n+m}+\frac{(-1)^{n+m}}{2}=0.
\end{eqnarray}
This is exactly the equation to derive UDD(1). In other words, our
$(n+m)$-pulse sequence has to be a first order sequence at least.
Now we make the first observation: Eq.~($20$) holds for any
combination of UDD($n$) and UDD($m$) with $n+m$ odd. Besides, parity
plays an important role in the following relations
\begin{eqnarray}
&&\sum_{j=1}^m\sum_{j^{''}<i^{'}}^n(-1)^{i+j}\sin\omega{t}(\delta_{j^{''}}-\delta_{i^{'}})\nonumber\\
&=&\sum_{i=1}^n\sum_{i^{'}>j^{''}}^m(-1)^{i+j}\sin\omega{t}(\delta_{(n+1-i)^{'}}-\delta_{(m+1-j)^{''}}),\nonumber\\
&=&\sum_{i=1}^n\sum_{i^{'}<j^{''}}^m(-1)^{i+j+m+n}\sin\omega{t}(\delta_{i^{'}}-\delta_{j^{''}}),
\end{eqnarray}
and
\begin{eqnarray}
&&\sum_{j=1}^m(-1)^{j+n+1}\sin\omega{t}(\delta_{j^{''}}-1)\nonumber\\
&=&-\sum_{j=1}^m(-1)^{j+n+1}\sin\omega{t}(\delta_{(m+1-j)^{''}}),\nonumber\\
&=&-\sum_{j=1}^m(-1)^{j+n+m}\sin\omega{t}(\delta_{j^{''}}).
\end{eqnarray}
As a result, if $n$ and $m$ have different parity, we have
\begin{eqnarray}
x_m(\omega{t})+z_m(\omega{t})+x_n(\omega{t})+z_n(\omega{t})=0.
\end{eqnarray}
Moreover, it is easy to verify\begin{eqnarray}
x_m(\omega{t})+z_m(\omega{t})-x_n(\omega{t})-z_n(\omega{t})=\Im(y_n(t)y_m^*(t)),\nonumber\\
\end{eqnarray}
since
\begin{eqnarray}
&&\sum_{i=1}^n\sum_{i^{'}<j^{''}}^m(-1)^{i+j}\sin\omega{t}(\delta_{i^{'}}-\delta_{j^{''}})\nonumber\\
&-&\sum_{j=1}^m\sum_{j^{''}<i^{'}}^n(-1)^{i+j}\sin\omega{t}(\delta_{j^{''}}-\delta_{i^{'}})\nonumber\\
&=&\sum_{i=1}^n\sum_{i^{'}<j^{''}}^m(-1)^{i+j}\sin\omega{t}(\delta_{i^{'}}-\delta_{j^{''}})\nonumber\\
&+&\sum_{j=1}^m\sum_{j^{''}<i^{'}}^n(-1)^{i+j}\sin\omega{t}(\delta_{i^{'}}-\delta_{j^{''}}),\nonumber\\
&=&\sum_{i=1}^n\sum_{j=1}^m(-1)^{i+j}\sin\omega{t}(\delta_{i^{'}}-\delta_{j^{''}}).
\end{eqnarray}
For $n+m$ is odd, solving Eq.~($23$) and Eq.~($24$) yields
\begin{eqnarray}
f_m(\omega{t})=-f_n(\omega{t})=\Im(y_n(t)y_m^*(t))/2.
\end{eqnarray}
Again the phase factors $\Delta_m(t)$ and $\Delta_n(t)$ are bounded
by the filter functions $y_m(t)$ and $y_n(t)$. Thus we have
completed the proof that both phase and coherence can be preserved
to the same order. In Fig.~\ref{fig1}, we give three examples of
independent UDD sequences with different parity, where
$f_m(\omega{t})$ and $f_n(\omega{t})$ are numerically calculated. It
can be seen that the phases are suppressed order by order through
increasing the pulse number of each UDD sequence.

\subsection{UDD Sequences with Same Parity}
\begin{figure}
\begin{center}
\scalebox{0.6}[0.6]{\includegraphics{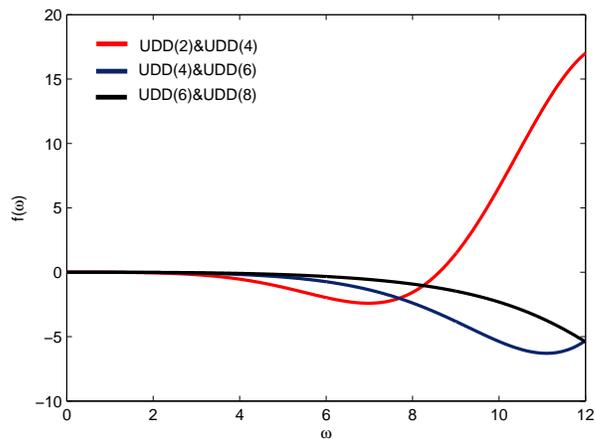}}\caption{The same
definition as Fig.~\ref{fig1}. Note that the solid lines completely
overlap the dashed counterparts as indicated by
Eq.~(27).{\label{fig2}}}
\end{center}
\end{figure}

We only consider two UDD sequences of even order, and two odd-order
sequences share essentially the same property because their middle
pulses do not flip the collective operator
$\sigma_{z_1}\sigma_{z_2}$. If $n$ and $m$ are even, from Eq.~($21$)
and Eq.~($22$) we know
\begin{eqnarray}
x_n(\omega{t})+z_n(\omega{t})=x_m(\omega{t})+z_m(\omega{t}).
\end{eqnarray}
A corollary is drawn: $y_n(t)y_m^*(t)$ is a real number if $n$ and
$m$ are even. Moreover, $x_m(\omega{t})+z_m(\omega{t})$ are not
bounded by filter functions anymore. As shown in Fig.~\ref{fig2},
phase evolutions are eliminated at a fixed level. Increasing pulse
number cannot improve the performance. Suppression of
$f_n(\omega{t})$ and $f_m(\omega{t})$ before $\omega{t}<2$ is due to
the fact: $\sin\omega{t}\approx\omega{t}$ if $\omega{t}$ is small.
As a consequence, $f_n(\omega{t})$ and $f_m(\omega{t})$ are
self-corrected to first order, see Eq.~($2$). While
$\sigma_{z_1}\sigma_{z_2}\omega{t}$ is effectively averaged at low
frequencies, the nonlinear part of
$\sigma_{z_1}\sigma_{z_2}\sin\omega{t}$ is uncontrollable.\par In
the present model the indirect coupling via a thermal bath is
commonly weak. If there exists highly nonlinear indirect coupling
between the qubits, we expect more distinctions will be observed by
using different UDD sequences.

\begin{figure}
\begin{center}
\scalebox{0.6}[0.5]{\includegraphics{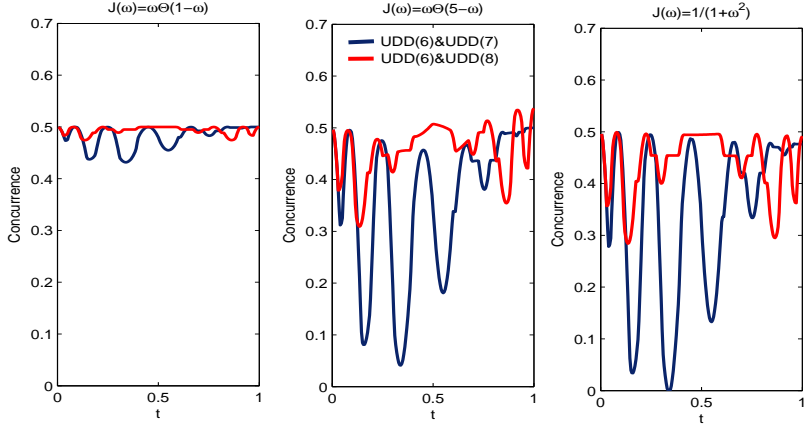}}\caption{Evolutions
of concurrence in three different environments. $\Theta(\cdot)$ is
Heaviside step function. In each of the three diagrams, red line
stands for the same parity, and the other colored line for different
parity.{\label{fig3}}}
\end{center}
\end{figure}

\subsection{Entanglement Dynamics}
In order to illustrate the difference caused by the parity,
especially the distinct oscillation patterns of quantum correlation
under different decoupling schemes, we numerically calculate the
entanglement dynamics. The initial bipartite state is chosen to be
$\frac{\sqrt2}{4}(|0\rangle+\sqrt3|1\rangle+\sqrt3|2\rangle+|3\rangle)$,
which is partially entangled. In Fig.~\ref{fig3} we have used
concurrence to measure the two-qubit entanglement \cite{ar31}.

We consider the dynamical evolution of concurrence versus time under
three kinds of environmental spectrum. We can see that for
$J(\omega)=\omega\Theta(1-\omega)$, the final concurrence at $t=1$
is perfectly kept at the initial level regardless of the parity.
This result is consistent with our earlier observation that the low
frequency part of $f(\omega{t})$ is self-corrected. However for the
other two spectrums, the UDD performances are obviously dependent on
the parity difference. For $J(\omega)=\omega\Theta(5-\omega)$, the
combination of UDD(6) and UDD(7) still preserves the initial
entanglement at the end of the decoupling cycle, while UDD(6) and
UDD(8) fails to do so. Since UDD(6) and UDD(8) are not able to
effectively suppress the indirect coupling between the two qubits,
the quantum correlation of their final state is much larger than the
initial one. Similar increase can be observed in soft-cutoff
spectrum ($J(\omega)=\frac{1}{1+\omega^2}$). Note that in this type
of spectrum concurrences cannot be well preserved in both
combinations owing to severe decoherence \cite{ar6,ar32}.

As shown in Fig.~\ref{fig3}, concurrences undergo violent
oscillations within the pulse intervals. In contrast with UDD(6) and
UDD(8), the concurrence that is controlled by the combination of
UDD(6) and UDD(7) drops more rapidly in the beginning. For
$J(\omega)=\frac{1}{1+\omega^2}$, the concurrence even reach zero at
one point, indicating the entanglement is completely lost. However
at the final stage of the decoupling period, the sequences with
different parity have a more smooth and steady concurrence, which
makes them more feasible since the loose timing constraint for
retrieving the final state, compared with the combination of UDD(6)
and UDD(8).

\section{Conclusion}
In this paper we give a detailed analysis of available DD sequences
which can be used to eliminate both qubit-environment and indirect
qubit-qubit coupling in a common bath. Exact dynamics under
arbitrary pulse sequences are derived. As we find out, it is
possible to apply UDD sequences independently on each of the two
qubit, and at the same time preserve the bipartite state to higher
order. This result greatly simplifies the scheme of universal NUDD
when dealing with correlated environments. We has proven that by
applying UDD($n$) and UDD($m$) with $n+m$ odd, the evolutions of all
density matrix elements are bounded by the single-qubit filter
functions of UDD($n$) and UDD($m$).\par In conclusion, we suggest
using UDD sequences with different parity to protect a dephasing
bipartite system, in case that the environments for individual qubit
are not completely independent. Besides, our results may find
applications in quantum information processing due to its superior
ability to dynamically switch off the interaction induced by a
common quantum bus.

\section{Acknowledgments}
Yu Pan wishes to thank Jiang-bin Gong for helpful comments on the
manuscript. This work is supported by National Nature Science
Foundation under Grant No. 60774099, No. 60821091, and No. 61134008.

\end{document}